\documentclass[twocolumn,showpacs,preprintnumbers,amsmath,amssymb]{revtex4}

\usepackage{graphicx}
\usepackage{dcolumn}
\usepackage{bm}

\begin{document}


\title{Comparative study of ordered and disordered Y$_{1-x}$Sr$_x$CoO$_{3-\delta}$}

\author{S. Fukushima}
 \email{fukush-s@sophia.ac.jp}

\author{T. Sato}
\author{D. Akahoshi}
\author{H. Kuwahara}
\affiliation{
Department of Physics, Sophia University\\
Chiyoda-ku, Tokyo 102-8554, JAPAN
}%



\begin{abstract}
We have succeeded in preparing $A$-site ordered- and disordered-Y$_{1/4}$Sr$_{3/4}$CoO$_{3-\delta}$ with various oxygen deficiencies $\delta$, and have made comparative study of their structural and physical properties.  
In the $A$-site ordered structure, oxygen vacancies order, and $\delta$ = 0.34 sample shows a weak ferromagnetic transition beyond 300 K.  
On the other hand, in the $A$-site disordered structure, no oxygen vacancy ordering is observed, and $\delta$ = 0.16 sample shows a ferromagnetic metallic transition around 150 K.  
$A$-site disordering destroys the orderings of oxygen-vacancies and orbitals, leading to the strong modification of the electronic phases.  
\end{abstract}

\maketitle

%
%

Cobalt oxides with perovskite-based structure exhibit many attracting phenomena such as ferromagnetism, metal-insulator transition, and spin-state transition.  
These phenomena are caused by the various spin states of Co ion such as low-spin (LS), intermediate-spin (IS), and high-spin (HS) states.  
La$_{1-x}$Sr$_{x}$CoO$_3$ is the most typical example:  
The ground state of $x$ = 0 compound is a nonmagnetic insulator with LS state of Co$^{3+}$.\cite{Senaris_JSSC_116,Asai_JPSJ_67}  
With increase of Sr content $x$, a ferromagnetic metallic state emerges in $x \ge$ 0.18, which is driven by the double-exchange interaction between Co$^{4+}$ in LS state and Co$^{3+}$ in IS state.\cite{Itoh_JPSJ_63,Senaris_JSSC_118}  

Among many perovskite cobaltites, Y$_{1-x}$Sr$_x$CoO$_{3-\delta}$ attracts our interest, because of its characteristic structural features and high magnetic-transition temperatures.  
$A$-site ordered-Y$_{1-x}$Sr$_x$CoO$_{3-\delta}$ has been first reported by Withers {\it et al}.\cite{Withers_JSSC_174} and Istomin {\it et al}.\cite{Istomin_Chem.Mater._15}  
This compound has large oxygen deficiency $\delta$.  
The structure of ordered-Y$_{1/4}$Sr$_{3/4}$CoO$_{2.66}$ can be regarded as the alternate stacking of $A$O and CoO$_{1.7}$ sheets, where $A$-site cations and oxygen vacancies concomitantly order as shown in Fig.~1(a).  
Y$_{1/4}$Sr$_{3/4}$O sheet stacks along the $c$-axis with translation $\langle 1/2 \text{, } 1/2 \text{, } 0\rangle$.  
CoO$_{1.3}$ and CoO$_2$ sheets alternately stack along the $c$-axis with translation $\langle 1/2 \text{, } 1/2 \text{, } 0\rangle$ to form the four times periodicity.  
For  $0.75 \le x \le 0.8$, $A$-site ordered-Y$_{1-x}$Sr$_x$CoO$_{3-\delta}$ undergoes a weak ferromagnetic transition at 335 K.\cite{Kobayashi_PRB_72}  
Ishiwata {\it et al}.~propose that the weak ferromagnetism is related to the orbital ordering.\cite{Ishiwata_PRB_75}   
In addition, the physical properties are sensitive to the variation of oxygen deficiency $\delta$.  
For example, in $x$ = 0.33, the ground state is transformed from an antiferromagnetic insulator to a ferromagnetic metal with a slight decrease of $\delta$.\cite{Maignan_JSSC_178}

\begin{figure}
\includegraphics[clip]{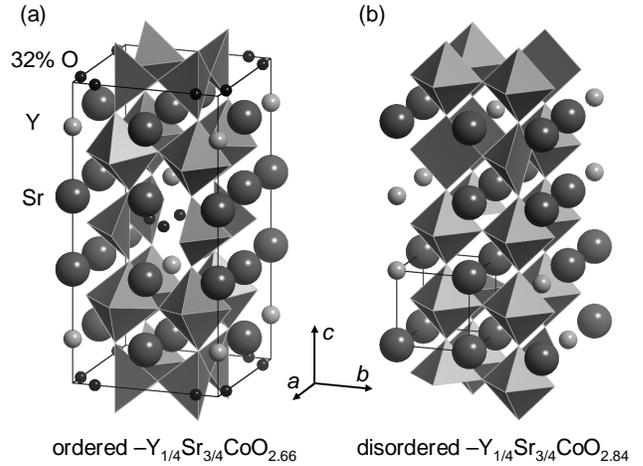}
\caption{\label{fig:structure} Schematic structure of (a) $A$-site ordered-Y$_{1/4}$Sr$_{3/4}$CoO$_{2.66}$ and (b) disordered-Y$_{1/4}$Sr$_{3/4}$CoO$_{2.84}$.  Small dark spheres in (a) represent oxygen site 30 \% occupied.  The ordered-Y$_{1/4}$Sr$_{3/4}$CoO$_{2.66}$ has CoO$_2$ layers at $z$ = 1/4 and 3/4, and CoO$_{1.3}$ layers at $z$ = 0 and 1/2 with translation $\langle 1/2 \text{, } 1/2 \text{, } 0\rangle$.  In disordered-Y$_{1/4}$Sr$_{3/4}$CoO$_{2.84}$, Y$^{3+}$ and Sr$^{2+}$ occupy $A$-site at random, and oxygen vacancies are randomly distributed.}
\end{figure}

The high magnetic transition temperature is due to the $A$-site ordered structure free from chemical disorder.  
The $A$-site arrangement in perovskite-based oxides plays an important role in determining their physical properties.\cite{Akahoshi_PRL_90}    
In $A$-site ordered-$R_{1/2}$Ba$_{1/2}$MnO$_3$ ($R$ = rare earth ion), ferromagnetic or charge-ordering transitions exceed room temperature, and a bicritical feature can be seen in the phase diagram.  
The random potential arising from $A$-site disordering suppresses long-range orders, and gives rise to the colossal magnetoresistive state near the bicritical region.  
In the cobalt oxide system just like the manganese oxide system, the $A$-site arrangement probably affects the physical properties through Co spin state and orbital arrangement.  
However, there are few reports on randomness effect of $A$-site ions in the cobalt oxides.\cite{Nakajima_JPSJ_74}  
In this study, we prepared $A$-site ordered- and disordered-Y$_{1/4}$Sr$_{3/4}$CoO$_{3-\delta}$ with various $\delta$ and have made a comparative study of them in order to investigate the $A$-site randomness effect on their structural and physical properties.

%
%

$A$-site ordered- and disordered-Y$_{1/4}$Sr$_{3/4}$CoO$_{3-\delta}$ were prepared in a polycrystalline form by solid state reaction.  
Mixed powders of Y$_2$O$_3$, SrCO$_3$, and CoO were heated at 1073~K in air with a few intermediate grindings, and sintered at 1423~K in air.  
Then the sample was annealed at 1173~K in Ar atmosphere.  
The resulting product has $A$-site ordered form.  
On the other hand, $A$-site disordered form is obtained by quenching the sample from 1473 to 77~K\@.   
We prepared ordered- and disordered-Y$_{1/4}$Sr$_{3/4}$CoO$_{3-\delta}$ with various $\delta$ through different annealing conditions.  
The detailed conditions will be described elsewhere.  
The values of $\delta$ were determined through iodometric titration with an accuracy of $\pm$ 0.01.  
The crystallographic analysis of the obtained samples was performed by X-ray-diffraction (XRD) method at room temperature.  
The resistivity was measured by a standard four-probe method from 5 to 350~K\@.   
The magnetic properties were measured using a Quantum Design, Physical Property Measurement System (PPMS) from 5 to 350~K\@.

%
%
 
From the XRD measurements, we confirmed that Y$_{1/4}$Sr$_{3/4}$CoO$_{2.66}$ annealed in Ar has the $A$-site ordered structure as shown in Fig.~1(a).  
$A$-site ordered-Y$_{1/4}$Sr$_{3/4}$CoO$_{3-\delta}$ with $\delta$ = 0.44 and 0.30 have the similar $A$-site ordered structure as that of $\delta$ = 0.34 sample, but the arrangement of oxygen vacancies are different from each other.  
On the other hand, quenched Y$_{1/4}$Sr$_{3/4}$CoO$_{3-\delta}$ ($0.15 \le \delta \le 0.27$) samples have a simple cubic perovskite structure.  
Bragg peaks arising from $A$-site and/or oxygen-vacancy orderings are not observed, indicating that Y$^{3+}$ and Sr$^{2+}$ randomly occupy $A$-site, and that oxygen vacancies are randomly distributed as shown in Fig.~1(b).  
These results indicate that $A$-site disordering makes the oxygen-vacancy order destabilized.  
The detailed results of the structural properties will be published elsewhere.

\begin{figure}
\includegraphics[clip]{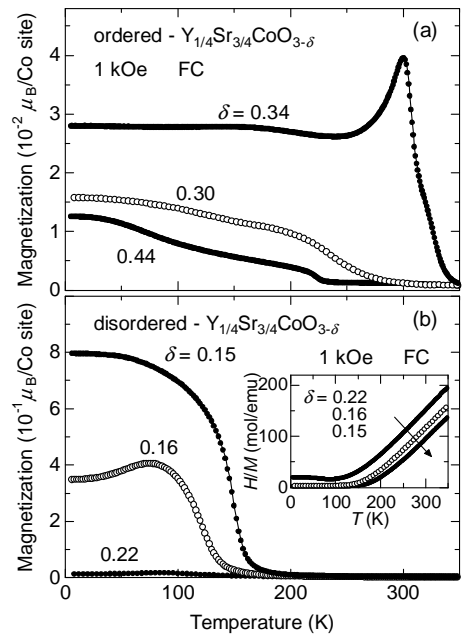}
\caption{\label{fig:MT} Temperature dependence of the magnetization of (a) $A$-site ordered-Y$_{1/4}$Sr$_{3/4}$CoO$_{3-\delta}$ ($\delta$ = 0.30, 0.34, and 0.44) and (b) disordered-Y$_{1/4}$Sr$_{3/4}$CoO$_{3-\delta}$ ($\delta$ = 0.15, 0.16, and 0.22).  The inset shows the inverse susceptibility of $A$-site disordered-Y$_{1/4}$Sr$_{3/4}$CoO$_{3-\delta}$.}
\end{figure}

Figure 2(a) shows the temperature dependence of the magnetization ($M$-$T$) of $A$-site ordered-Y$_{1/4}$Sr$_{3/4}$CoO$_{3-\delta}$.  
The magnetization of  $\delta$ = 0.34 shows an abrupt increase below 330~K\@.  
The onset of the magnetization corresponds to the weak ferromagnetic transition.  
The magnetic field dependence of the magnetization ($M$-$H$) shows a ferromagnetic behavior at 300 K, but the magnetization remains unsaturated even at 80 kOe ($M$ = 0.19 $\mu_{\rm B}/{\rm Co}$).  
This magnetic behavior is consistent with that reported by Kobayashi {\it et al}.\cite{Kobayashi_PRB_72}.  
On the basis of structural refinement, they propose two models for the magnetic order.  
One is a canted antiferromagnetic state, and the other is a ferrimagnetic state due to the antiferromagnetic coupling between the HS and IS states of Co$^{3+}$.\cite{Ishiwata_PRB_75}  
Slight increase or decrease of $\delta$ from 0.34 suppresses the weak ferromagnetic phase around 325~K\@.  
The magnetization of $\delta$ = 0.30 gradually increases below 250~K, and shows weak ferromagnetic behavior at low temperatures.  
On the other hand, $\delta$ = 0.44 shows a magnetic transition around 230~K\@.  
The magnetization of $\delta$ = 0.44 sample linearly depends on applied magnetic fields at 5~K, while the $M$-$H$ curves of $\delta$ = 0.34 and 0.30 indicate that they have ferromagnetic components.  
With decrease of $\delta$ from 0.44, that is, with increase of Co valence, the ferromagnetic component tends to increase at low temperatures.  
The magnetization of $\delta$ = 0.44, 0.34, and 0.30 at 80 kOe are 0.127, 0.132, and 0.23 $\mu_{\rm B}/{\rm Co}$, respectively.  

Figure 2(b) shows the $M$-$T$ curves of $A$-site disordered-Y$_{1/4}$Sr$_{3/4}$CoO$_{3-\delta}$.  
The magnetization of $\delta$ = 0.15 exhibits an abrupt jump around 150~K\@.  
The $M$-$H$ curve of $\delta$ = 0.15 shows the typical ferromagnetic hysteresis below 150 K, in contrast to that of ordered-Y$_{1/4}$Sr$_{3/4}$CoO$_{2.66}$.  
The saturation magnetization of $\delta$ = 0.15 at 5~K is 1.2 $\mu_{\rm B}/{\rm Co}$.
With increase of $\delta$, the ferromagnetic phase is drastically suppressed.  
The magnetic transition temperatures of $A$-site disordered-Y$_{1/4}$Sr$_{3/4}$CoO$_{3-\delta}$ are much lower than those of $A$-site ordered-Y$_{1/4}$Sr$_{3/4}$CoO$_{3-\delta}$.  
As shown in the inset of Fig.~2(b), the temperature dependence of inverse susceptibility ($H/M$-$T$) of disordered-Y$_{1/4}$Sr$_{3/4}$CoO$_{3-\delta}$ ($\delta$ = 0.15, 0.16, and 0.22) obeys Curie-Weiss law above 250 K with Weiss temperatures $\theta$ = 190.3, 169.1, and 134.1~K, and effective moments estimated from the observed slopes in this region are $P_{\rm eff}$ = 3.06, 3.02, and 3.00 $\mu_{\rm B}/{\rm Co}$, respectively.  
This result indicates that magnetic interaction among Co-spins is ferromagnetic for disordered-Y$_{1/4}$Sr$_{3/4}$CoO$_{3-\delta}$ with $\delta$ = 0.15 to 0.22.

\begin{figure}
\includegraphics[clip]{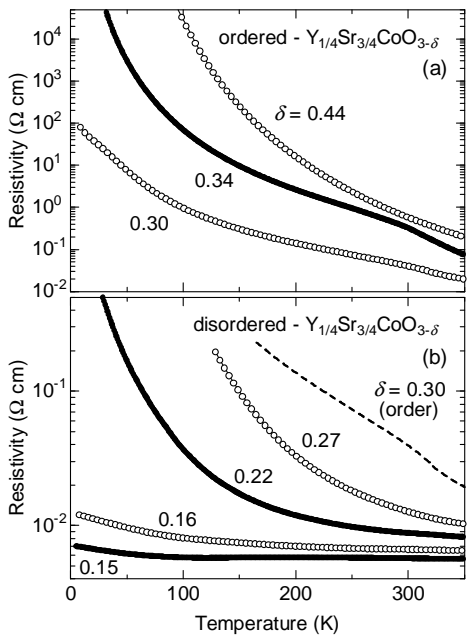}
\caption{\label{fig:rT} Temperature dependence of the resistivity of (a) $A$-site ordered-Y$_{1/4}$Sr$_{3/4}$CoO$_{3-\delta}$ ($\delta=$0.30, 0.34, and 0.44) and (b) disordered-Y$_{1/4}$Sr$_{3/4}$CoO$_{3-\delta}$ ($\delta=$0.15, 0.16, 0.22, and 0.27). The resistivity of ordered-Y$_{1/4}$Sr$_{3/4}$CoO$_{2.70}$ is also shown (broken line) for comparison.}  
\end{figure}

Figure 3(a) shows the temperature dependence of the resistivity ($\rho$-$T$) of $A$-site ordered-Y$_{1/4}$Sr$_{3/4}$CoO$_{3-\delta}$.  
The resistivities of all the samples show semiconducting or insulating behavior.  
With decrease of $\delta$, the resistivity abruptly drops over the whole temperatures.  
Taking into account the results of the $M$-$T$ curves, this implies the increase of ferromagnetic metallic components except for $\delta$ = 0.34 having large ferromagnetic component.  
$\delta$ = 0.34 shows a clear kink around 300 K, which corresponds to the enhanced weak ferromagnetic transition.  
Figure 3(b) shows the $\rho$-$T$ curves of $A$-site disordered-Y$_{1/4}$Sr$_{3/4}$CoO$_{3-\delta}$.  
$\delta$ = 0.27 shows insulating behavior.  
With decrease of $\delta$, the resistivity is steeply decreased similar to the case of $A$-site ordered-Y$_{1/4}$Sr$_{3/4}$CoO$_{3-\delta}$, and $\delta$ = 0.15 sample shows almost metallic behavior.  

Then, we will discuss the effect of $A$-site disordering on the structural properties of Y$_{1/4}$Sr$_{3/4}$CoO$_{3-\delta}$.  
In the $A$-site ordered structure ($0.30 \le \delta \le 0.44 $), the periodic-potential and lattice-distortion associated with Y/Sr ordering are likely to stabilize the oxygen-vacancy order and orbital order.  
In contrast, in $A$-site disordered-Y$_{1/4}$Sr$_{3/4}$CoO$_{3-\delta}$ ($0.15 \le \delta \le 0.27 $), the random-potential and lattice-distortion due to Y/Sr disordering destroy the oxygen-vacancy order.  

Such randomness strongly affects the physical properties as well as the structural properties.  
As shown in Fig.~2, the magnetic transition temperatures of $A$-site disordered-Y$_{1/4}$Sr$_{3/4}$CoO$_{3-\delta}$ are much lower than those of ordered-Y$_{1/4}$Sr$_{3/4}$CoO$_{3-\delta}$.  
Furthermore, a deviation of $\delta$ from 0.34 as well as $A$-site disordering has much effect on the magnetic structure.  
In $A$-site ordered-Y$_{1/4}$Sr$_{3/4}$CoO$_{3-\delta}$, $\delta$ = 0.34 shows the highest magnetic transition temperature (the weak ferromagnetic transition at 330~K).  
As described above, the oxygen-vacancy ordered structure depends on $\delta$.  
The oxygen-vacancy ordering pattern near $\delta$ = 0.34 may be essential for the weak ferromagnetism.  
The relation between the arrangement of oxygen-vacancy and the magnetic structure is now under investigation.  
The magnetic and transport properties of $A$-site disordered-Y$_{1/4}$Sr$_{3/4}$CoO$_{3-\delta}$ resemble those of La$_{1-x}$Sr$_{x}$CoO$_3$.\cite{Itoh_JPSJ_63}  
In La$_{1-x}$Sr$_{x}$CoO$_3$, the magnetization and the magnetic transition temperature increase continuously with increasing Sr content $x$ or Co valence, and the ferromagnetic metallic state is dominant for $x \ge 0.18$.  
The ferromagnetic metallic state of $A$-site disordered-Y$_{1/4}$Sr$_{3/4}$CoO$_{3-\delta}$ is likely to be attributed to double-exchange interaction just like the case of La$_{1-x}$Sr$_{x}$CoO$_3$.\cite{Itoh_JPSJ_63}

%
%

In summary, we have prepared $A$-site ordered- and disordered-Y$_{1/4}$Sr$_{3/4}$CoO$_{3-\delta}$, and have investigated their structural and physical properties.  
Oxygen vacancy orders in the $A$-site ordered structure, but not in the $A$-site disordered structure.  
$A$-site disordered-Y$_{1/4}$Sr$_{3/4}$CoO$_{3-\delta}$ exhibits the magnetic transition below 150~K, which is quite lower than those of ordered-Y$_{1/4}$Sr$_{3/4}$CoO$_{3-\delta}$.  
The $A$-site randomness suppresses magnetic order in cobalt oxide Y$_{1/4}$Sr$_{3/4}$CoO$_{3-\delta}$ as reported in manganese oxide $R_{1/2}$Ba$_{1/2}$MnO$_3$.\cite{Akahoshi_PRL_90}

%
%

This work was supported by Iketani Science and Technology Foundation, the Matsuda Foundation, the Asahi Glass Foundation, and by Grant-in-Aid for Scientific Research on Priority Areas (No.451) from the Ministry of Education, Culture, Sports, Science and Technology (MEXT), Japan.


\end{document}